\documentclass[nofootinbib,aps,showpacs,groupedaddress,twocolumn,10pt,floatfix]{revtex4}

\newcommand{\ifproofpre}[2]{#1}

\usepackage{graphicx}
\usepackage{amsmath}
\usepackage{amssymb}
\usepackage{mathtools}
\usepackage{isotope}
\usepackage{wignert}

\newcommand{\MeV}{{\mathrm{MeV}}}

\newcommand{\fm}{{\mathrm{fm}}}

\newcommand{\Nmax}{{N_\text{max}}}

\newcommand{\hw}{{\hbar\omega}}

\newcommand{\Einf}{{E_\infty}}  
\newcommand{\kinf}{{k_\infty}}  
\newcommand{\rinf}{r_\infty}  

\newcommand{\scrN}{{\mathcal{N}}}  

\begin{document}

\title{Natural orbital description of the halo nucleus \boldmath$\isotope[6]{He}$}

\author{Chrysovalantis Constantinou}
\altaffiliation[Present address: ]{Center for Theoretical Physics, Sloane Physics Laboratory,
Yale University, New Haven, Connecticut 06520-8120, USA.}
\author{Mark A. Caprio}
\affiliation{Department of Physics, University of Notre Dame, Notre Dame, Indiana 46556-5670, USA}

\author{James P. Vary}
\author{Pieter Maris}
\affiliation{Department of Physics and Astronomy, Iowa State University, Ames, Iowa 50011-3160, USA}

\date{\today}

\begin{abstract}
\textit{Ab initio} calculations of nuclei face the challenge of
simultaneously describing strong short-range internucleon correlations
and the long-range properties of weakly-bound halo nucleons.  Natural
orbitals, which diagonalize the one-body density matrix, provide a
basis which is better matched to the physical structure of the
many-body wave function.  We demonstrate that the use of natural
orbitals significantly improves convergence for \textit{ab initio}
no-core configuration interaction calculations of the neutron halo
nucleus $\isotope[6]{He}$, relative to the traditional oscillator
basis.
\end{abstract}


\maketitle

\section{Introduction}
\label{sec-intro}

\textit{Ab initio} calculations of nuclear
structure~\cite{pieper2004:gfmc-a6-8,neff2004:cluster-fmd,hagen2007:coupled-cluster-benchmark,quaglioni2009:ncsm-rgm,bacca2012:6he-hyperspherical,shimizu2012:mcsm,dytrych2013:su3ncsm,barrett2013:ncsm,baroni2013:7he-ncsmc}
face the challenge of describing a complex, multiscale quantum
many-body system.  The goal is to directly solve the many-body problem
for a system of protons and neutrons, with realistic internucleon
interactions~\cite{wiringa1995:nn-av18,entem2003:chiral-nn-potl,shirokov2007:nn-jisp16,epelbaum2009:nuclear-forces}.
However, the nucleus is governed by a
strong, short-range interaction.  Short-range correlations, tightly-bound $\alpha$
clusters~\cite{freer2007:cluster-structures}, and weakly-bound halo
nucleons~\cite{jonson2004:light-dripline,tanihata2013:halo-expt}
introduce dynamics over differing length scales and energy scales,
within the same nucleus, which must be simultaneously described within
the same many-body calculation.

Natural
orbitals~\cite{loewdin1955:natural-orbitals-part1,shull1955:natural-orbitals-helium,loewdin1956:natural-orbital,davidson1972:natural-orbital,mahaux1991:single-particle,stoitsov1993:natural-orbital-correlation}
provide a means of adapting the single-particle basis to better match
the physical structure of the many-body wave function.  Natural
orbitals are obtained by diagonalizing the one-body density matrix,
deduced from a preliminary many-body calculation using an
initial reference single-particle basis.  The Laguerre function
basis~\cite{helgaker2000:electron-structure,mccoy2016:lgalg} has been
used as the starting point for natural orbitals in atomic
electron-structure
calculations~\cite{shull1955:natural-orbitals-helium}, while we start
from the harmonic oscillator orbitals~\cite{moshinsky1996:oscillator}
more familiar to the nuclear structure
context~\cite{suhonen2007:nucleons-nucleus}.  The natural orbital
basis builds in important contributions from high-lying orbitals of
the initial basis~--- for the present application, high-lying
oscillator shells~--- thereby accelerating the convergence of wave
functions, energies, and other observables.

In this work, we present a framework for \textit{ab initio} no-core
configuration interaction (NCCI)~\cite{barrett2013:ncsm} calculations
with a natural orbital basis and demonstrate improved convergence for
the lightest neutron halo
nucleus~$\isotope[6]{He}$~\cite{jonson2004:light-dripline}.  When used
with recently-proposed infrared (IR) basis-extrapolation
schemes~\cite{furnstahl2012:ho-extrapolation,more2013:ir-extrapolation},
we show that natural orbitals provide improved independence of basis
parameters for predictions of energy and radius observables.

\section{Natural orbitals}

\begin{figure}[tp]
\begin{center}
\includegraphics[width=\ifproofpre{0.9}{0.5}\hsize]{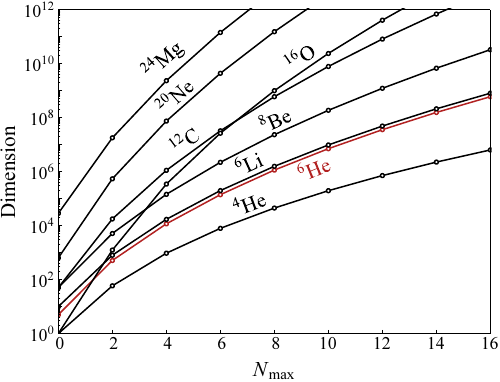}
\end{center}
\vspace{-2em}
\caption{Growth of the NCCI problem dimension as a function of the number
  of oscillator excitations $\Nmax$ included in the basis, for
  selected nuclides, including $\isotope[6]{He}$~(red curve).
The dimensions
  shown are
  for spaces with zero angular momentum projection ($M=0$) and positive parity.  
}
\label{fig-dimension}
\end{figure}

In NCCI calculations, the nuclear many-body Schr\"odinger equation is
formulated as a matrix eigenproblem, where the Hamiltonian is
represented within a basis of Slater determinants, \textit{i.e.},
antisymmetrized products of single-particle states.  Conventionally,
harmonic oscillator orbitals~\cite{moshinsky1996:oscillator} are used,
and the basis is truncated to a maximum allowed number $\Nmax$ of
oscillator excitations~\cite{barrett2013:ncsm}.  The calculated wave
functions, energies, and observables depend upon both the truncation
$\Nmax$ and the oscillator length $b$ of the basis (or, equivalently,
the oscillator energy $\hw\propto b^{-2}$).  The solution of the full,
untruncated many-body problem could, in principle, be obtained to any
desired accuracy, by retaining a sufficiently complete basis set.
However, the dimension of the NCCI problem increases rapidly with the
number of nucleons and included single-particle excitations, as shown
in Fig.~\ref{fig-dimension}.  Currently available computational
resources therefore limit the convergence of calculated states and
observables~\cite{maris2009:ncfc,maris2013:ncsm-pshell,caprio2015:berotor-ijmpe}.

We therefore seek a physically-adapted basis, in which the nuclear
many-body wave function can be efficiently and accurately described,
subject to the constraint of accessible problem dimensions.  The
\textit{natural orbital} basis minimizes the mean occupation of states
above the Fermi
surface~\cite{schaefer1971:brueckner-self-consistency,mahaux1991:single-particle},
thus reducing the contribution of high-lying orbitals in describing
the many-body wave function.  Intuitively, the natural orbitals may be
understood as representing an attempt to recover the basis in which
the many-body wave function most resembles a single Slater
determinant.  Although we cannot expect to reduce the complex,
highly-correlated nuclear wave function to a single Slater
determinant, as assumed in the Hartree-Fock approximation, we may
expect that transforming to a more natural single-particle basis could
enhance the role of a comparatively small set of dominant Slater
determinants, and thereby accelerate convergence within a Slater
determinant expansion.

An NCCI state of good total angular momentum can only, in general, be
obtained as a superposition of several Slater determinants of $nljm$
single particle states (here, $n$ is the radial quantum number, $l$
labels the orbital angular momentum, $j$ labels the resultant angular
momentum after coupling to the spin, and $m$ labels its projection).
Therefore, we cannot, in general, expect the nuclear eigenfunctions to
resemble a single Slater determinant.  Rather, we hope to recover wave
functions which most resemble a single \textit{configuration} of
nucleons over $nlj$ orbitals.  For this purpose, we consider the
\textit{scalar densities}
$\rho^{(0)}_{ab}\equiv\tme{\Psi}{[c^\dagger_b\tilde{c}_a]_{00}}{\Psi}$,
where $c^\dagger_a$ represents the creation operator for a nucleon in
orbital $a=(n_al_aj_a)$ and the brackets $[\cdots]_{00}$ represent
spherical tensor coupling to zero angular momentum.  For a single
configuration $\tket{\Psi}$, the scalar densities are diagonal, and
the diagonal entries give the occupations of the contributing orbitals
[$\tbracket{\scrN_a}=(2j_a+1)^{1/2}\rho^{(0)}_{aa}$].\footnote{That is, the total occupation number operator for an orbital is
                  $\scrN_a\equiv\sum_{m_a} c^\dagger_{a,m_a}
                  c_{a,m_a}=\pm(2j_a+1)^{1/2}[c^\dagger_a\tilde{c}_a]_{00}$,
                  where the sign is to be taken  according to the choice of conjugation phase convention
                  $\tilde{c}_{a,m_a}\equiv(-)^{j_a\pm m_a}c_{a,-m_a}$.}
Otherwise, for
the general case of a many-body state $\tket{\Psi}$, natural orbitals
are defined by diagonalizing this scalar density matrix.  The scalar
density matrix only connects orbitals of the same $l$ and $j$,
\textit{i.e.}, differing at most in their radial quantum number $n$,
so the transformation to natural orbitals induces a change of basis on
the radial functions separately within each $lj$ space [$\tket{n'lj} =
  \sum_n a^{(lj)}_{n',n} \tket{nlj}$].

An initial NCCI calculation is carried out in the oscillator basis.
This provides a scalar density matrix for the ground-state wave
function, which is then diagonalized to yield the natural orbitals.
The eigenvalue associated with a natural orbital represents its mean
occupation in the many-body wave function.  We order the natural
orbitals by decreasing eigenvalue of the density
matrix~\cite{mahaux1991:single-particle}, \textit{i.e.}, starting with
$n=0$ for the natural orbital with highest eigenvalue
[$\tbracket{\scrN_{0lj}}\geq\tbracket{\scrN_{1lj}}\geq\cdots$],
thereby providing an $n$ quantum number for an $\Nmax$-type truncation
scheme (see Ref.~\cite{caprio2012:csbasis}).
\begin{figure}[tp]
\begin{center}
\includegraphics[width=\ifproofpre{0.9}{0.5}\hsize]{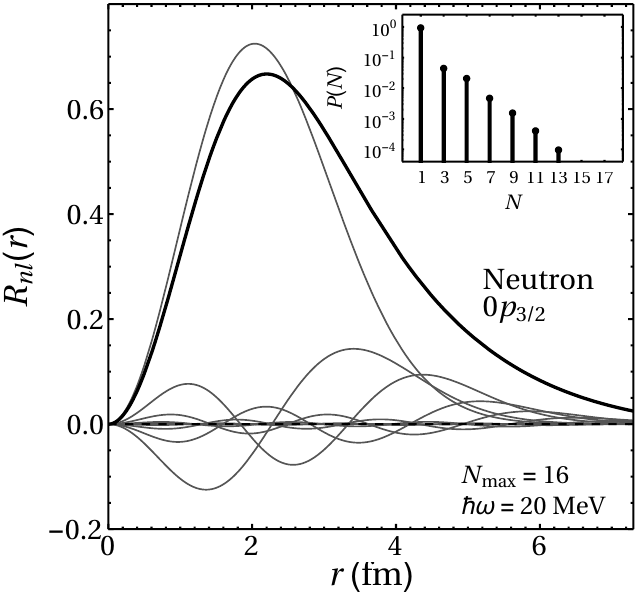}
\end{center}
\vspace{-2em}
\caption{Radial wave function for the neutron $0p_{3/2}$ natural
  orbital (heavy curve) derived from the $\isotope[6]{He}$ ground
  state calculation in the harmonic oscillator basis, along with the
  contributions from individual oscillator basis functions (gray
  curves).  The squared amplitudes $P(N)$ of these contributions are
  shown in the inset. The initial oscillator basis for this
  calculation has $\Nmax=16$ and $\hw=20\,\MeV$.}
\label{fig-radial}
\end{figure}

The lowest $p_{3/2}$ natural orbital obtained from the
$\isotope[6]{He}$ ground-state one-body densities is illustrated in
Fig.~\ref{fig-radial}, taking an example from the NCCI calculations
presented below.  In a traditional shell-model description,
$\isotope[6]{He}$ consists of two protons and two neutrons in a filled
$s$ shell ($N\equiv2n+l=0$), plus two neutrons in the valence $p$
shell ($N=1$).  In the oscillator-basis NCCI calculations, this
general structure is reflected in a nearly-filled $s$ shell.  The next
most heavily occupied orbital is then the neutron $0p_{3/2}$.  We
observe that the corresponding natural
orbital [Fig.~\ref{fig-radial}] receives extended contributions from high-lying oscillator
shells and thus acquires a substantial large-$r$ tail compared to the
oscillator orbital, as may be expected for a weakly-bound halo
nucleon.

The NCCI calculation using the natural orbital basis is no
more computationally difficult than the original oscillator basis
calculation.  It is simply necessary, in preparation, to carry out a
similarity transformation of the input Hamiltonian, as described in
Refs.~\cite{hagen2006:gdm-realistic,caprio2012:csbasis,caprio2014:cshalo}.

\section{Results}
\label{sec-results}

Several experimental properties of the ground state of
$\isotope[6]{He}$ support the interpretation that it consists of a
weakly-bound two-neutron halo surrounding a tightly-bound $\alpha$
core~\cite{jonson2004:light-dripline,tanihata2013:halo-expt}.  The two-neutron
separation energy for $\isotope[6]{He}$ is only $0.97\,\MeV$, out of a
total binding energy of $29.27\,\MeV$~\cite{npa2002:005-007}.
Experimentally, the onset of halo structure along the $\isotope{He}$
isotopic chain is indicated by a jump in the measured charge and
matter radii, from $\isotope[4]{He}$ to $\isotope[6]{He}$.  The root
mean square (RMS) point-proton distribution radius $r_p$, which may be
deduced~\cite{friar1997:charge-radius-correction} from the measured
charge radius, increases by $\sim32\%$ from $\isotope[4]{He}$
[$r_p=1.462(6)\,\fm$] to $\isotope[6]{He}$
[$r_p=1.934(9)\,\fm$]~\cite{wang2004:6he-radius-laser,brodeur2012:6he-8he-mass,lu2013:laser-neutron-rich}.
This increase may be understood as a consequence of halo structure,
arising from the recoil of the charged $\alpha$ core against the halo
neutrons (as well as possible contributions from swelling of the
$\alpha$ core~\cite{lu2013:laser-neutron-rich}).

The initial oscillator basis NCCI calculations from which we derive
natural orbitals for $\isotope[6]{He}$ cover a range of oscillator
basis parameters $\hw=10\,\MeV$ to $40\,\MeV$ with truncations
$\Nmax\leq16$, as considered in Ref.~\cite{caprio2014:cshalo}.  The
NCCI calculations are carried out using the code
MFDn~\cite{maris2010:ncsm-mfdn-iccs10,aktulga2013:mfdn-scalability},
with the JISP16 two-body internucleon
interaction~\cite{shirokov2007:nn-jisp16} plus Coulomb interaction.
\begin{figure}[tp]
\begin{center}
\includegraphics[width=\ifproofpre{0.9}{0.5}\hsize]{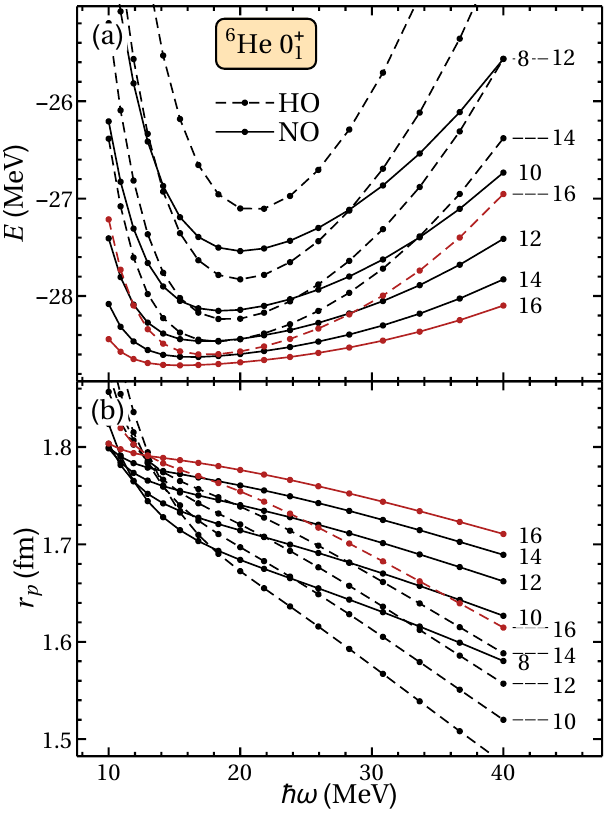}
\end{center}
\vspace{-2em}
\caption{Comparison of $\isotope[6]{He}$ ground state properties
  calculated using harmonic oscillator (HO, dashed lines) and natural
  orbital (NO, solid lines) bases: (a)~energy and (b)~ground state proton
  radii.  These are shown as functions of the oscillator basis $\hw$,
  for $\Nmax=10$ to $16$ (as labeled).}
\label{fig-scan}
\end{figure}

The ground state energy eigenvalues obtained for $\isotope[6]{He}$
using the harmonic oscillator (dashed lines) and natural orbital 
(solid lines) bases are compared in Fig.~\ref{fig-scan}(a). The energies
from the natural orbital calculations are lower (thus, by the
variational principle, closer to the true value) than those from the
harmonic oscillator calculations and are also less dependent
upon $\hw$.  The improvement in convergence afforded by the natural
orbitals ranges from approximately one step in $\Nmax$ in the vicinity
of the variational minimum ($\hw\approx15\text{--}20\,\MeV$) to
several steps in $\Nmax$ towards the ends of the calculated $\hw$
range.\footnote{We consider steps of $2$ in the number of oscillator quanta,
since we restrict
our attention to the positive parity sector for the present calculations.}

For the $\isotope[6]{He}$ proton radius, shown in Fig.~\ref{fig-scan}(b), the natural orbital basis NCCI
calculations lead the harmonic oscillator basis calculations in convergence
by more than one step in $\Nmax$ at $\hw\approx20\,\MeV$ and by
several steps in $\Nmax$ at the high end of the $\hw$ range.
These radii obtained from the natural orbital calculations are also less dependent
upon $\hw$
than those obtained from the harmonic oscillator calculations.  

The goal we set out to achieve is to find the true results for
observables as they would be obtained in the full,
infinite-dimensional space. Although full convergence is not achieved,
even with the natural orbitals, the improved convergence motivates us
to attempt to obtain estimates of the converged results via basis extrapolation
methods~\cite{bogner2008:ncsm-converg-2N,maris2009:ncfc,coon2012:nscm-ho-regulator,furnstahl2012:ho-extrapolation}.  Infrared extrapolation
schemes~\cite{furnstahl2012:ho-extrapolation,more2013:ir-extrapolation,furnstahl2014:ir-expansion,furnstahl2015:ir-extrapolation-cc-16o,wendt2015:ir-extrapolations}
are based on the premise that the solution of the many-body problem in
a truncated space effectively imposes infrared (long-range) and
ultraviolet (short-range) cutoffs. For bases with high enough $\hw$
(and $\Nmax$) ultraviolet convergence is assumed, and any remaining
incomplete convergence is attributed to the failure of the basis to
reproduce the long-range tail of the many-body wave function.

A basis consisting of harmonic oscillator orbitals with no more than
$N$ quanta cannot fully resolve long-range physics beyond the
classical turning point
$L(N,\hw)=[2(N+3/2)]^{1/2}\,b(\hw)$~\cite{furnstahl2012:ho-extrapolation},\footnote{Specifically, we adopt
the cutoff $L_2(N,\hw)=[2(N+2 + 3/2)]^{1/2}\,b(\hw)$ from
Ref.~\cite{furnstahl2012:ho-extrapolation} and take
$N=\Nmax+1$, since this is the highest number of oscillator quanta
accessible to the neutrons in $\isotope[6]{He}$.  
The single-particle space spanned by the natural
orbitals is identical to that of the underlying harmonic oscillator
orbitals, so the estimated length cutoff remains unchanged.}
 where 
$b(\hw)=(\hbar c)/[(m_Nc^2)(\hw)]^{1/2}$ is again the oscillator length, with $m_N$ the nucleon mass.
The calculated energy and observables are expected to depend only on
this cutoff $L$, approaching the true converged values as
$L\rightarrow\infty$.
For energy eigenvalues, it is expected that~\cite{furnstahl2014:ir-expansion,furnstahl2015:ir-extrapolation-cc-16o}
\begin{equation}
\label{eqn-E-ir}
E(L) = \Einf +  a_0 e^{-2 \kinf L},
\end{equation}
where $\Einf$, $a_0$, and $\kinf$ are to be deduced as fitting
parameters from the results of calculations in truncated spaces.
Taking $L\rightarrow\infty$, we extract $\Einf$ as an estimate for the
true energy.  For mean square radii, it is expected that, letting
$\beta\equiv2\kinf L$,
\begin{equation}
r^{2}(L) = \rinf^2 [1 - (c_0+c_1\beta^{-2})\beta^3 e^{-\beta}],
\label{eqn-rsqr-ir}
\end{equation}
for $\beta\gg1$, where $\rinf$, $c_0$, and $c_1$ are similarly 
deduced from calculations in truncated spaces, and
$\rinf$ provides an estimate of the true RMS radius.
\begin{figure}[tp]
\begin{center}
\includegraphics[width=\ifproofpre{1}{0.9}\hsize]{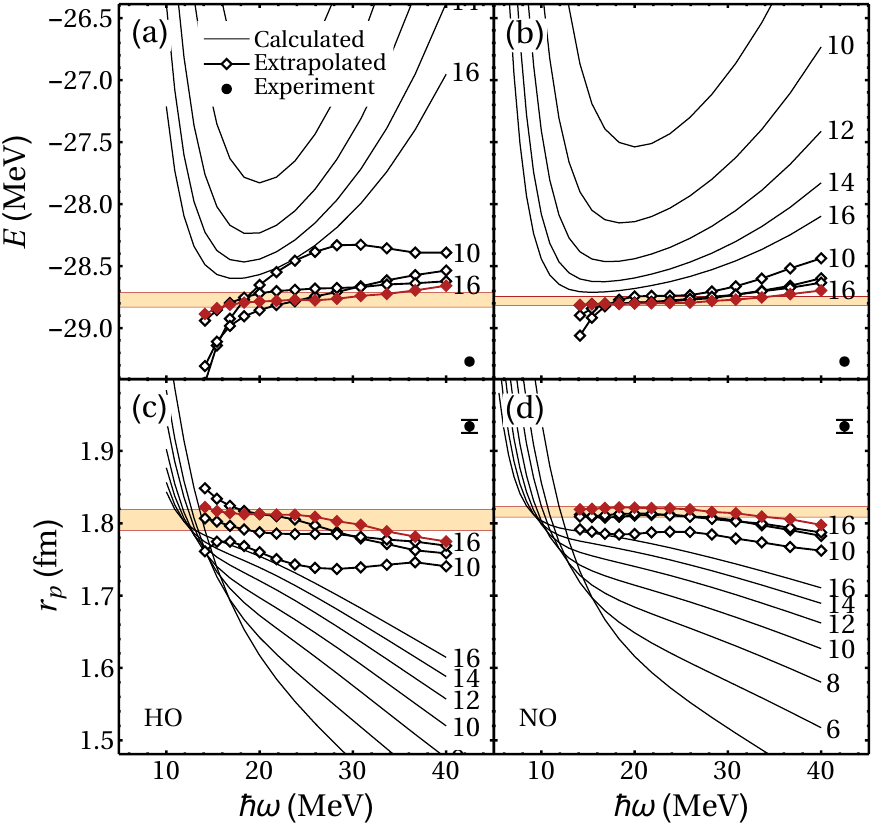}
\end{center}
\vspace{-2em}
\caption{Infrared basis extrapolations for the $\isotope[6]{He}$
  ground state energy~(top) and point proton radius~(bottom), based on
  calculations in the harmonic oscillator basis~(left) and natural
  orbital basis~(right).  The extrapolations (diamonds) are shown
  along with the underlying calculated results (plain lines) as
  functions of $\hw$ at fixed $\Nmax$ (as indicated).  Experimental
  values (circles) are shown with uncertainties.  The shaded bands
  reflect the mean values and standard deviations of the extrapolated
  results, at the highest $\Nmax$, over the $\hw$ range considered.}
\label{fig-extrap}
\end{figure}

The extrapolated values for the $\isotope[6]{He}$ ground state energy
and proton radius are shown in Fig.~\ref{fig-extrap}.  We restrict
ourselves to a straightforward application of~(\ref{eqn-E-ir})
and~(\ref{eqn-rsqr-ir}), based on three-point extrapolation in $\Nmax$
at fixed $\hw$.  Calculations at low $\hw$ may not provide the assumed
ultraviolet convergence, while poor infrared convergence at high $\hw$
leads to an excessively large correction and thus poor
extrapolation.

The extrapolated $\isotope[6]{He}$ ground state energies from
the natural orbital NCCI calculations [Fig.~\ref{fig-extrap}(b)] are
considerably less $\hw$-dependent than the extrapolated
energies from the harmonic oscillator NCCI calculations
[Fig.~\ref{fig-extrap}(a)].  The extrapolations obtained for different
$\Nmax$ are also considerably more consistent (in the figure, $\Nmax$
refers to the highest $\Nmax$ in the three-point
extrapolation).  
The extrapolated ground state energies obtained with the harmonic
oscillator and natural orbital bases at $\hw=20\,\MeV$ (chosen close to the variational energy minimum) and
$\Nmax=16$ are consistent with each other to within their respective
variations, giving $E\approx-28.79\,\MeV$ and $E\approx-28.80\,\MeV$,
respectively.

Once the many-body calculation is under control, any remaining
deviation of calculated values from nature may be attributed to
deficiencies in the internucleon interaction.  Comparing to the
experimental binding energy of $29.27\,\MeV$ thus indicates that the
JISP16 interaction underbinds $\isotope[6]{He}$ by
$\sim0.5\,\MeV$.\footnote{The present extrapolations for the $\isotope[6]{He}$ ground state energy
are consistent with
the estimate
   $E=-28.8(1)\,\MeV$~\cite{maris2013:ncsm-pshell} obtained from the
  \textit{ad hoc}
  exponential basis extrapolation
  scheme for the oscillator basis~\cite{bogner2008:ncsm-converg-2N,maris2009:ncfc}.
}  (For comparison, the binding
of $\isotope[4]{He}$ obtained with JISP16 matches experiment to within
$\sim0.003\,\MeV$~\cite{maris2009:ncfc}.)

The extrapolated proton radii extracted from the NCCI calculations
with the natural orbital basis [Fig.~\ref{fig-extrap}(d)] similarly
demonstrate a reduced $\hw$ dependence and $\Nmax$ dependence, as
compared to the extrapolations from the oscillator-basis calculations
[Fig.~\ref{fig-extrap}(c)].  At the highest calculated $\Nmax$
($\Nmax=16$), the extrapolated $r_p$ varies only by $\sim0.02\,\fm$
across the range of $\hw$ values shown ($\hw\approx14\,\MeV$ to
$40\,\MeV$), and the $\Nmax$ dependence is comparable.
We must emphasize that the variations in extrapolated values at best
provide a rough guide to how well we can trust these extrapolated
values as reflecting the true radius which would be obtained in an
untruncated many-body calculation.  Nonetheless, the
$\hw$-independence and $\Nmax$-independence of the calculations at the
$\sim0.02\,\fm$ level is reassuring.

Taking the extrapolated proton
radius at $\hw=20\,\MeV$ and $\Nmax=16$ as representative gives
$r_p\approx1.82\,\fm$.\footnote{The present extrapolated result for the  
$\isotope[6]{He}$ proton radius is consistent with previous
estimates~\cite{caprio2014:cshalo} based on the ``crossover point''~\cite{bogner2008:ncsm-converg-2N} of successive $\Nmax$
curves in a plot such as Fig.~\ref{fig-scan}(b).}  Thus, it would appear that the \textit{ab
  initio} NCCI calculations with the JISP16 interaction, while
qualitatively reproducing the increase in proton radius with the onset
of halo structure in $\isotope[6]{He}$, do yield a quantitative
shortfall of $\sim0.12\,\fm$ (or $\sim6\%$) for the proton
radius of $\isotope[6]{He}$.

\section{Conclusion}
\label{sec-conclusion} 

Describing the nuclear many-body wave function within truncated
spaces is challenging due to the need to
describe, simultaneously, long-range asymptotics and short-range
correlations.  Natural orbitals, obtained here by diagonalizing
one-body density matrices from initial NCCI calculations using the
harmonic oscillator basis, build in contributions from high-lying
oscillator shells, thereby accelerating convergence.  

In the present application to the halo nucleus $\isotope[6]{He}$,
improvement is by about one step in $\Nmax$ near the variational
minimum in $\hw$, and significantly more for other $\hw$ values
(Fig.~\ref{fig-scan}). To put these gains in perspective, we note that
an increment in $\Nmax$ results in an increase in matrix dimension of
about a factor of $3$--$5$, as seen in Fig.~\ref{fig-dimension}, with
much larger increase in the computational
costs~\cite{vary2009:ncsm-mfdn-scidac09}.  Although full convergence is still not achieved, the
calculations using natural orbitals provide improved
basis parameter independence for extrapolations with respect to the infrared cutoff of the
basis (Fig.~\ref{fig-extrap}).

The successful application of natural orbitals to \textit{ab initio}
nuclear NCCI calculations presented here provides a starting point for
exploring ideas (some taken from electron structure theory) which may more fully realize the potential of the
natural orbital approach:

(1)~NCCI calculations based on natural orbitals
yield improved one-body densities which can, in turn, be diagonalized
to yield new natural orbitals.  Natural orbitals constructed through such an iterative
method can rapidly build in additional contributions from high-lying
shells, thereby potentially further accelerating convergence~\cite{bender1966:natural-orbital-iterative-hydride,davidson1972:natural-orbital}.

(2)~An improved reference basis for the initial NCCI calculation
may also boost the convergence of the subsequent natural orbital calculations. For instance, the Laguerre
functions, commonly used as the starting point for natural
orbitals in electron-structure calculations, also have the correct
exponential asymptotics for nucleons bound by a
finite-range potential.

(3)~The structure of nuclear excited states
can vary markedly from that of the ground state.  Natural orbitals
constructed by diagonalizing the density matrices from excited states,
rather than from the ground state,
may more effectively accelerate convergence of those excited
states~\cite{davidson1972:natural-orbital}.

(4)~Finally, natural orbitals are conducive to a more efficient
many-body truncation scheme than the conventional oscillator $\Nmax$
scheme.  The eigenvalues associated with the natural orbitals, by
providing an estimate of the mean occupation of each orbital in the
many-body wave function, also suggest a means of estimating the
relative importance of Slater determinants involving these orbitals.

\section*{Acknowledgements}
\label{sec-acknowledgements} 

We thank G.~Hupin for valuable discussions on the formulation of the
nuclear natural orbital problem and M.~A.~McNanna for carrying out
informative preliminary studies in one dimension.  This material is based upon work supported by the
U.S.~Department of Energy, Office of Science, under Award
Numbers~DE-FG02-95ER-40934, DESC0008485 (SciDAC/NUCLEI), and
DE-FG02-87ER40371.  This research used computational resources of
the University of Notre Dame Center for Research Computing and of the
National Energy Research Scientific Computing Center (NERSC), a
U.S.~Department of Energy, Office of Science, user facility supported
under Contract~DE-AC02-05CH11231.

\providecommand{\APSLONG}{}
\providecommand{\ELSEVIER}{}


\clearpage


\end{document}